\documentstyle[twocolumn,eqsecnum,aps]{revtex}

\newcommand{\nc}{\newcommand}
\nc{\rnc}{\renewcommand}
\nc{\bc}{\begin{center}}
\nc{\ec}{\end{center}}
\nc{\e}[1]{{\em#1\/}}
\rnc{\b}[1]{{\bf#1}}
\nc{\comment}[1]{}
\nc{\spos}[2]{\makebox(0,0)[#1]{${\scriptstyle#2}$}}
\nc{\be}{\begin{equation}}
\nc{\ee}{\end{equation}}
\nc{\bea}{\begin{eqnarray}}
\nc{\eea}{\end{eqnarray}}
\nc{\disp}{\displaystyle}
\nc{\ade}{\mbox{$A$-$D$-$E$}}
\nc{\calN}{{\cal N}}
\nc{\calC}{{\cal C}}
\nc{\calM}{{\cal M}}
\nc{\calS}{{\cal S}}
\nc{\phit}{\hat{\varphi}}
\nc{\chit}{\hat{\chi}}
\nc{\hcalN}{\hat{\calN}}
\nc{\hcalS}{\hat{\calS}}
\nc{\hS}{\tilde{S}}             
\nc{\sigmad}{\sigma^\dagger}
\nc{\psid}{\psi^\dagger}
\def\smat#1{\mbox{\scriptsize{\mbox{$\pmatrix{#1}$}}}}
\def\hV{\hat{V}}
\def\hN{\hat{N}}

\font\tenmsb=msbm10
\font\sevenmsb=msbm7
\font\fivemsb=msbm5
\newfam\msbfam
\textfont\msbfam=\tenmsb
\scriptfont\msbfam=\sevenmsb
\scriptscriptfont\msbfam=\fivemsb
\def\Bbb#1{{\fam\msbfam\relax#1}}

\begin{document}
\draft
\preprint{HEP/123-qed}
\title{Integrable Boundaries, Conformal Boundary Conditions\\
and $A$-$D$-$E$ Fusion Rules
}
\author{Roger E. Behrend and Paul A. Pearce}
\address{
Department of Mathematics and Statistics\\
University of Melbourne, Parkville, Victoria 3052, Australia
}
\author{Jean-Bernard Zuber}
\address{
Service de Physique Th\'eorique\\
CEA-Saclay, 91191 Gif-sur-Yvette,
Cedex, France
}
\date{\today}
\maketitle
\widetext
\begin{abstract}
The $sl(2)$ minimal theories are
labelled by a Lie algebra pair $(A,G)$ where $G$ is of
$A$-$D$-$E$ type. For these theories on a cylinder we conjecture a complete
set of
conformal boundary conditions labelled by the nodes of the tensor product
graph $A\otimes G$.
The cylinder partition functions are given by fusion rules arising from the
graph fusion algebra of $A\otimes G$. We further conjecture that, for
each conformal boundary condition, an integrable boundary condition exists
as a solution of the boundary
Yang-Baxter equation for the associated lattice model. The theory is
illustrated using the $(A_4,D_4)$
or 3-state Potts model. 
\end{abstract}

\narrowtext

\section{INTRODUCTION}
\label{sec:level1}

The study of conformal boundary conditions~\cite{Cardy89} continues to be
an active area of research
with applications in statistical mechanics and string theory. The problem of
a general classification of conformal boundary conditions has seen a
revival of interest  recently. For theories 
with a diagonal torus partition function
it is known that there is a conformal boundary condition associated to each
operator appearing in the
theory. Moreover, the fusion rules of these boundary operators
are just given by the bulk fusion algebra and thus 
by the Verlinde formula~\cite{Verlinde88}. In contrast, for non-diagonal
theories, the fusion rules are not
known in general and  it is not even known 
what constitutes a complete set of conformal boundary
conditions. 
Indeed, these questions have only been
resolved~\cite{AfflOS98,FuchsS98} very recently for the simplest
non-diagonal theory, namely, the
critical 3-state Potts model. In this letter we conjecture a complete set
of conformal boundary
conditions, fusion rules and cylinder partition functions for the $sl(2)$
minimal models.

The $sl(2)$ minimal models in the bulk are
classified~\cite{CapItzZub87} by a pair of simply laced Dynkin diagrams
$(A,G)$ of  type
\be
(A,G)=\cases{
(A_{h-1},A_{g-1})&\cr
(A_{h-1},D_{(g+2)/2}),\quad g\ \mbox{even}&\cr
(A_{h-1},E_6), \quad g=12 &\cr
(A_{h-1},E_7), \quad g=18&\cr
(A_{h-1},E_8), \quad g=30.}
\ee
Here $h$ and $g$ are the coprime Coxeter numbers of
$A$ and $G$ and the central charges are
\be
c=1-\disp{6(h-g)^2\over hg}.
\ee
\goodbreak

$$\mbox{}$$
\vspace{1.0in}
\mbox{}

We conjecture that for these theories a complete set of conformal
boundary conditions $i$ and the corresponding boundary operators $\phit_i$ are
labelled by $i\in(A,G)$
\be
\phit_i:\quad i=(r,a)\in (A,G)
\label{boundops}
\ee
where $r$, $a$ are nodes on the Dynkin diagram of $A$ and $G$ respectively. We
will use $G$ to denote the Dynkin diagram and the adjacency
matrix of this graph. We use $r, r_1, r_2$ to denote nodes of $A_{h-1}$;
$s, s_1, s_2$ for the nodes of
$A_{g-1}$; $a, a_1, a_2, b$ for the nodes of $G$ and $i,j$
to label nodes in the pair $(A,G)$.

We now introduce fused adjacency matrices (intertwiners) and graph
fusion matrices.
The fused adjacency matrices
$V_s$ with $s=1,\ldots,g-1$ are defined recursively by the
$sl(2)$ fusion algebra
\be
V_s=V_2 V_{s-1} -V_{s-2}
\ee
subject to the initial conditions $V_1=I$ and $V_2=G$. 
The matrices $V_s$ 
are symmetric and mutually commuting with entries given by a
Verlinde-type formula
\be
V_{sa}{}^b=(V_s)_a{}^b
=\sum_{m\in \mbox{\scriptsize Exp}(G)}{\hS_{sm}\over
\hS_{1m}}\;\Psi_{am}\Psi_{bm}^*.
\ee
where the columns of the unitary matrices $\hS$ and $\Psi$ are
the eigenvectors of
the adjacency matrices $A_{g-1}$ and $G$ respectively and the sum is over
the Coxeter exponents of $G$
with multiplicities.
We assume the graph $G$ has a distinguished endpoint node labelled $a=1$
such that
$\Psi_{1m}>0$ for all $m$.
This is at least the case for \ade\ graphs.
In this notation we define the fundamental intertwiner as
 $\hV_s{}^a=V_{s1}{}^a$.

The graph fusion matrices $\hat{N}_a$ with $a\in G$
 were introduced by Pasquier~\cite{Pas}.
These  are defined
by the  Verlinde-type formula~\cite{DFZPZ}
\be
\hN_{ab}{}^c=(\hat{N}_a)_b{}^c= \sum_{m\in \mbox{\scriptsize Exp}(G)}
{\Psi_{am} \Psi_{bm} \Psi^*_{cm}\over \Psi_{1m}},\qquad a,b,c\in G.
\ee
These matrices satisfy the matrix recursion relation
\be
G\hN_a=\sum_{b\in G}G_a{}^b\hN_b
\ee
and initial conditions $\hN_1=I$ and $\hN_2=G$ where 2 denotes the unique
node adjacent to 1.  The numbers $\hN_{ab}{}^c$ are
the structure constants 
of the graph fusion algebra
\be
\hN_a\hN_b=\sum_{c\in G}\hN_{ab}{}^c \hN_c.
\ee
All the entries of the fused
adjacency matrices $V_s$ 
are nonnegative integers.
For a proper choice of the eigenvectors  and of the
node $1$, the entries of the graph fusion matrices $\hN_a$
are also integers,
and with the exception of 
$D_{2n+1}$
and $E_7$, they are nonnegative.
A key identity relating the fused adjacency matrices and graph fusion
matrices is
\be
V_s \hN_a =\sum_{b\in G} V_{sa}{}^b \hN_b\ .
\label{keyid}
\ee

\section{Fusion Rules}

Let $i_1
$, $i_2 
$ and $i_3
\in (A,G)$ and
consider the tensor product graph
$A\otimes G$ with
distinguished node $i=1$ given by $i=(r,a)=(1,1)$. Then we conjecture
that the fusion rules for the boundary operators (\ref{boundops}) are
\be
\phit_{i_1}\;\times\;\phit_{i_2}=\sum_{i_3\in
(A,G)}\calN_{i_1 i_2}{}^{i_3}\,\phit_{i_3}
\ee
where $\calN_{i_1}$ are just the graph fusion matrices associated with the
tensor product graph $A\otimes G$
\be
\calN_{i_1 i_2}{}^{i_3}=
\calN_{(r_1,a_1)(r_2,a_2)}{}^{(r_3,a_3)}=N_{r_1 r_2}{}^{r_3}
\hN_{a_1 a_2}{}^{a_3}
\ee
where $N_{r_1}$ are the graph fusion matrices for $A_{h-1}$.
Let $\varphi_{r,s}$ 
be the primary chiral fields
with respect to the Virasoro algebra. Then the operators
$\phit_i=\phit_{r,a}$ are related to
$\varphi_{r,s}$ by the intertwining relation
\be
\sum_{b\in G} \phit_{r,b}\,(\hV^T \hV)_b{}^a
= \sum_{s\in A_{g-1}}\varphi_{r,s} \hV_s{}^a \label{opintertw}
\ee
where $\hV$ is the fundamental adjacency matrix intertwiner
defined in sec. I. By equality in (\ref{opintertw})  we mean that
the operators on either side satisfy the same algebra
under fusion.

We define a conjugation operator $C(a)=a^*$ to be the identity
except for $D_{4n}$ graphs where the eigenvectors
 $\Psi_{am}$ are complex and conjugation
corresponds to the
${\Bbb Z}_2$ Dynkin diagram automorphism. It then follows that
 $\hN_{a^*b}{}^c=\hN_{ca}{}^b$.
We conjecture that the coefficients of
the cylinder partition functions $Z_{i_1|i_2}$ of the $sl(2)$ minimal
theories are given by the
fusion product $\phit_{i_1}^\dagger\times\phit_{i_2}$, that is
\begin{mathletters}\label{CPFs}
\be
Z_{i_1|i_2}(q)=\sum_{i_3\in (A,G)} \calN_{i_1^* i_2}{}^{i_3}\;\chit_{i_3}(q).
\ee
More explicitly,
\bea
&&Z_{(r_1,a_1)|(r_2,a_2)}(q)\nonumber\\
&=&\sum_{(r_3,a_3)\in(A_{h-1},G)}
\calN_{(r_1,a_1^*) (r_2,a_2)}{}^{(r_3,a_3)}\; \chit_{r_3,a_3}(q)\label{cpf1}\\
&=&
\sum_{(r,s)\in (A_{h-1},A_{g-1})}\chi_{r,s}(q)  N_{r r_1}{}^{r_2}
V_{sa_1}{}^{a_2}
\label{cpf2}
\eea
\end{mathletters}
where, in terms of Virasoro characters,
\be
\chit_{r,a}(q)
=\sum_{s\in A_{g-1}}\chi_{r,s}(q) \hV_s{}^a\ .
\ee
The equivalence of the two forms (\ref{cpf1}) and (\ref{cpf2}) of the
cylinder partition functions
follows from the identity (\ref{keyid})
with $a=1$. The result (\ref{CPFs}) is not
entirely new but generalizes
and encompasses several previous results~\cite{SaleurB89,Cardy89,PasqS90}.
Note that the matrices $N_r\otimes V_s$ form a representation of the
fusion algebra of the minimal model.

\section{Critical 3-state Potts}

As an example we consider the $\calM(A_4,D_4)$ or critical 3-state Potts
model. To avoid redundancy, we consider the folded
$(T_2,D_4)$ model as shown graphically in Figure~1.

The complete list~\cite{AfflOS98,FuchsS98} of conformal boundary
conditions, conjugate fields $\phit$ and
associated characters $\chit$ is
\bea
\begin{array}{rclrclcl}
A&=&(1,1)=(4,1)&\phit_{1,1}&=&I&\ &\chi_0+\chi_3\\
B&=&(1,3)=(4,3)&\phit_{1,3}&=&\psi&&\chi_{2/3}\\
C&=&(1,4)=(4,4)&\phit_{1,4}&=&\psi^\dagger&&\chi_{2/3}\\
BC&=&(2,1)=(3,1)&\phit_{2,1}&=&\epsilon&&\chi_{2/5}+\chi_{7/5}\\
AC&=&(2,3)=(3,3)&\phit_{2,3}&=&\sigma&&\chi_{1/15}\\
AB&=&(2,4)=(3,4)&\phit_{2,4}&=&\sigma^\dagger&&\chi_{1/15}\\
F&=&(1,2)=(4,2)\quad&\phit_{1,2}&=&\eta&&\chi_{1/8}+\chi_{13/8}\\
N&=&(2,2)=(3,2)&\phit_{2,2}&=&\xi&&\chi_{1/40}\!+\!\chi_{21/40}\nonumber
\end{array}
\eea

The fused adjacency matrices of $G=D_4$ are
\bea
V_1=V_5&=&\smat{1&0&0&0\cr 0&1&0&0\cr 0&0&1&0\cr 0&0&0&1},\quad
V_2=V_4=\smat{0&1&0&0\cr 1&0&1&1\cr 0&1&0&0\cr 0&1&0&0},\nonumber\\
V_3&=&\smat{0&0&1&1\cr 0&2&0&0\cr 1&0&0&1\cr 1&0&1&0}.
\eea

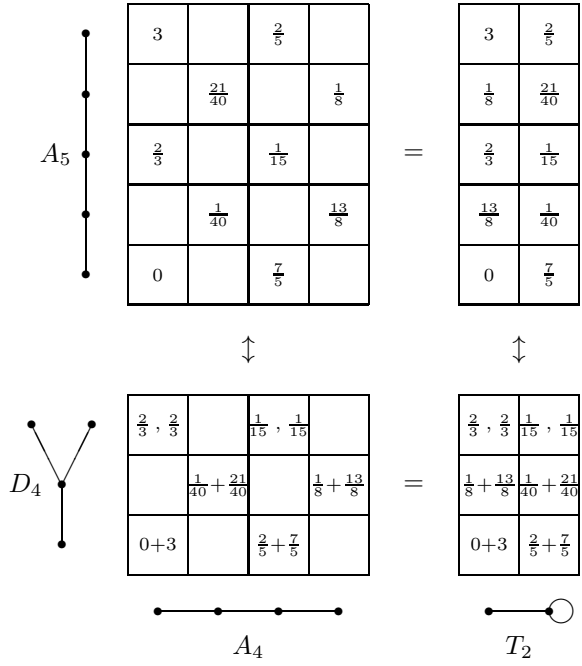
\begin{figure}[htb]
\nc{\sm}[1]{{\scriptstyle #1}}
\nc{\pos}[2]{\makebox(0,0)[#1]{$#2$}}
\setlength{\unitlength}{8mm}
\begin{center}
\begin{picture}(9.5,11)
\put(2,1.5){\begin{picture}(4,3)
\multiput(0,0)(1,0){5}{\line(0,1){3}}\multiput(0,0)(0,1){4}{\line(1,0){4}}
\put(0.5,0.5){\spos{}{0+3}}\put(2.5,0.5){\spos{}{\frac{2}{5}+\frac{7}{5}}}
\put(1.5,1.5){\spos{}{\frac{1}{40}+\frac{21}{40}}}\put(3.5,1.5){\spos{}{\frac{1}
{8}+\frac{13}{8}}}
\put(0.5,2.5){\spos{}{\frac{2}{3}\:,\:\frac{2}{3}}}\put(2.5,2.5){\spos{}{\frac{1
}{15}\:,\:\frac{1}{15}}}
\end{picture}}
\put(7.5,1.5){\begin{picture}(2,3)
\multiput(0,0)(1,0){3}{\line(0,1){3}}\multiput(0,0)(0,1){4}{\line(1,0){2}}
\put(0.5,0.5){\spos{}{0+3}}\put(1.5,0.5){\spos{}{\frac{2}{5}+\frac{7}{5}}}
\put(0.5,1.5){\spos{}{\frac{1}{8}+\frac{13}{8}}}\put(1.5,1.5){\spos{}{\frac{1}{4
0}+\frac{21}{40}}}
\put(0.5,2.5){\spos{}{\frac{2}{3}\:,\:\frac{2}{3}}}\put(1.5,2.5){\spos{}{\frac{1
}{15}\:,\:\frac{1}{15}}}
\end{picture}}
\put(2,6){\begin{picture}(4,5)
\multiput(0,0)(1,0){5}{\line(0,1){5}}\multiput(0,0)(0,1){6}{\line(1,0){4}}
\put(0.5,0.5){\spos{}{0}}\put(2.5,0.5){\spos{}{\frac{7}{5}}}
\put(1.5,1.5){\spos{}{\frac{1}{40}}}\put(3.5,1.5){\spos{}{\frac{13}{8}}}
\put(0.5,2.5){\spos{}{\frac{2}{3}}}\put(2.5,2.5){\spos{}{\frac{1}{15}}}
\put(1.5,3.5){\spos{}{\frac{21}{40}}}\put(3.5,3.5){\spos{}{\frac{1}{8}}}
\put(0.5,4.5){\spos{}{3}}\put(2.5,4.5){\spos{}{\frac{2}{5}}}
\end{picture}}
\put(7.5,6){\begin{picture}(2,5)
\multiput(0,0)(1,0){3}{\line(0,1){5}}\multiput(0,0)(0,1){6}{\line(1,0){2}}
\put(0.5,0.5){\spos{}{0}}\put(1.5,0.5){\spos{}{\frac{7}{5}}}
\put(0.5,1.5){\spos{}{\frac{13}{8}}}\put(1.5,1.5){\spos{}{\frac{1}{40}}}
\put(0.5,2.5){\spos{}{\frac{2}{3}}}\put(1.5,2.5){\spos{}{\frac{1}{15}}}
\put(0.5,3.5){\spos{}{\frac{1}{8}}}\put(1.5,3.5){\spos{}{\frac{21}{40}}}
\put(0.5,4.5){\spos{}{3}}\put(1.5,4.5){\spos{}{\frac{2}{5}}}
\end{picture}}
\put(2.5,0.9){\begin{picture}(3,0)\put(0,0){\line(1,0){3}}\multiput(0,0)(1,0){4}
{\spos{}{\bullet}}\end{picture}}
\put(8,0.9){\begin{picture}(1.4,0)\put(0,0){\line(1,0){1}}\multiput(0,0)(1,0){2}
{\spos{}{\bullet}}\put(1.2,0){\circle{0.4}}\end{picture}}
\put(0.9,2){\begin{picture}(0,2)\put(0,0){\line(0,1){1}}\multiput(0,0)(0,1){2}
{\spos
{}{\bullet}}
\put(0,1){\line(1,2){0.5}}\put(0,1){\line(-1,2){0.5}}\multiput(-0.5,2)(1,0){2}
{\spos{}{\bullet}}\end{picture}}
\put(1.3,6.5){\begin{picture}(0,4)\put(0,0){\line(0,1){4}}\multiput(0,0)(0,1){5}
{\spos{}{\bullet}}\end{picture}}
\put(4,0.3){\pos{}{A_4}}\put(8.5,0.3){\pos{}{T_2}}
\put(0.3,3){\pos{}{D_4}}\put(0.8,8.5){\pos{}{A_5}}
\put(6.75,3){\pos{}{=}}\put(6.75,8.5){\pos{}{=}}
\put(4,5.25){\pos{}{\updownarrow}}\put(8.5,5.25){\pos{}{\updownarrow}}
\end{picture}
\end{center}
\caption{Folding and orbifold duality relating the tensor product graph
$T_2\otimes D_4$ to
$A_4\otimes D_4$ and $A_4\otimes A_5$. The conformal weights of the 8
conformal boundary conditions
of the
3-state Potts model appear in the boxes of the $T_2\otimes D_4$ theory.}
\end{figure}

\noindent
The unitary matrix which diagonalizes $D_4$ is
\be
\Psi ={1\over\sqrt{3}}
\mbox{\smat{
{1\over\sqrt{2}}&{1\over\sqrt{2}}&1&1\cr
\sqrt{3\over 2}&-\sqrt{3\over 2}&0&0\cr
{1\over\sqrt{2}}&{1\over\sqrt{2}}&\omega&\omega^2\cr
{1\over\sqrt{2}}&{1\over\sqrt{2}}&\omega^2&\omega
}}
\ee
where $\omega=\exp(2\pi i/3)$ is a primitive cube root of unity.
The graph fusion matrices of $D_4$ are
\bea
&&\hat{N}_1=\smat{1&0&0&0\cr 0&1&0&0\cr 0&0&1&0\cr 0&0&0&1},\quad
\hat{N}_2=\smat{0&1&0&0\cr 1&0&1&1\cr 0&1&0&0\cr 0&1&0&0},\nonumber\\
&&\hat{N}_3=\smat{0&0&1&0\cr 0&1&0&0\cr 0&0&0&1\cr 1&0&0&0},\quad
\hat{N}_4=\smat{0&0&0&1\cr 0&1&0&0\cr 1&0&0&0\cr 0&0&1&0} \ .
\eea
The graph fusion matrices of $T_2$ are
\be
N_1=N^1=\mbox{\smat{1&0\cr 0&1}},\quad N_2=N^2=\mbox{\smat{0&1\cr 1&1}}.
\ee
The intertwiner $\hV$ and  conjugation $C$ are
\be
\hV=\mbox{\smat{1&0&0&0\cr 0&1&0&0\cr 0&0&1&1\cr 0&1&0&0\cr 1&0&0&0}},\qquad
C=\mbox{\smat{1&0&0&0\cr 0&1&0&0\cr 0&0&0&1\cr 0&0&1&0}}.
\ee
The  conjugation operator $C$ acts on the right to raise and lower
indices in the fusion
matrices $\hat{N}^a=\hat{N}_a C$.

The complete fusion rules of boundary fields are given as follows:
\bea
&&\smat{
I&\epsilon&\eta&\xi&\psi&\sigma&\psid&\sigmad\cr
\epsilon&\epsilon^2&\epsilon\eta&\epsilon\xi&\epsilon\psi&\epsilon\sigma&
\epsilon\psid&\epsilon\sigmad\cr
\eta&\eta\epsilon&\eta^2&\eta\xi&\eta\psi&\eta\sigma&\eta\psid&\eta\sigmad\cr
\xi&\xi\epsilon&\xi\eta&\xi^2&\xi\psi&\xi\sigma&\xi\psid&\xi\sigmad\cr
\psi&\psi\epsilon&\psi\eta&\psi\xi&\psi^2&\psi\sigma&\psi\psid&\psi\sigmad\cr
\sigma&\sigma\epsilon&\sigma\eta&\sigma\xi&\sigma\psi&\sigma^2&\sigma\psid&
\sigma\sigmad\cr
\psid&\psid\epsilon&\psid\eta&\psid\xi&\psid\psi&\psid\sigma&{\psid}^2&\psid\sigmad\cr
\sigmad&\sigmad\epsilon&\sigmad\eta&
\sigmad\xi&\sigmad\psi&\sigmad\sigma&\sigmad
\psid&{\sigmad}^2\cr }\nonumber\\
&=&\sum_{r=1}^2 \sum_{a=1}^4 N^r\otimes \hat{N}^a\; \phit_{r,a}\nonumber\\
&=&\!\!\smat{1&0&0&0&0&0&0&0\cr
						   0&1&0&0&0&0&0&0\cr
         0&0&1&0&0&0&0&0\cr
         0&0&0&1&0&0&0&0\cr
         0&0&0&0&0&0&1&0\cr
         0&0&0&0&0&0&0&1\cr
         0&0&0&0&1&0&0&0\cr
         0&0&0&0&0&1&0&0}I\!+\!\!
   \smat{0&1&0&0&0&0&0&0\cr
						   1&1&0&0&0&0&0&0\cr
         0&0&0&1&0&0&0&0\cr
         0&0&1&1&0&0&0&0\cr
         0&0&0&0&0&0&0&1\cr
         0&0&0&0&0&0&1&1\cr
         0&0&0&0&0&1&0&0\cr
         0&0&0&0&1&1&0&0}\epsilon\nonumber\\
&+&\!\!\smat{0&0&1&0&0&0&0&0\cr
						   0&0&0&1&0&0&0&0\cr
         1&0&0&0&1&0&1&0\cr
         0&1&0&0&0&1&0&1\cr
         0&0&1&0&0&0&0&0\cr
         0&0&0&1&0&0&0&0\cr
         0&0&1&0&0&0&0&0\cr
         0&0&0&1&0&0&0&0}\eta\!+\!\!
   \smat{0&0&0&1&0&0&0&0\cr
						   0&0&1&1&0&0&0&0\cr
         0&1&0&0&0&1&0&1\cr
         1&1&0&0&1&1&1&1\cr
         0&0&0&1&0&0&0&0\cr
         0&0&1&1&0&0&0&0\cr
         0&0&0&1&0&0&0&0\cr
         0&0&1&1&0&0&0&0}\xi\nonumber\\
&+&\!\!\smat{0&0&0&0&1&0&0&0\cr
						   0&0&0&0&0&1&0&0\cr
         0&0&1&0&0&0&0&0\cr
         0&0&0&1&0&0&0&0\cr
         1&0&0&0&0&0&0&0\cr
         0&1&0&0&0&0&0&0\cr
         0&0&0&0&0&0&1&0\cr
         0&0&0&0&0&0&0&1}\psi\!+\!\!
   \smat{0&0&0&0&0&1&0&0\cr
						   0&0&0&0&1&1&0&0\cr
         0&0&0&1&0&0&0&0\cr
         0&0&1&1&0&0&0&0\cr
         0&1&0&0&0&0&0&0\cr
         1&1&0&0&0&0&0&0\cr
         0&0&0&0&0&0&0&1\cr
         0&0&0&0&0&0&1&1}\sigma\nonumber\\
&+&\!\!ß\smat{0&0&0&0&0&0&1&0\cr
						   0&0&0&0&0&0&0&1\cr
         0&0&1&0&0&0&0&0\cr
         0&0&0&1&0&0&0&0\cr
         0&0&0&0&1&0&0&0\cr
         0&0&0&0&0&1&0&0\cr
         1&0&0&0&0&0&0&0\cr
         0&1&0&0&0&0&0&0}\psid\!+\!\!
   \smat{0&0&0&0&0&0&0&1\cr
						   0&0&0&0&0&0&1&1\cr
         0&0&0&1&0&0&0&0\cr
         0&0&1&1&0&0&0&0\cr
         0&0&0&0&0&1&0&0\cr
         0&0&0&0&1&1&0&0\cr
         0&1&0&0&0&0&0&0\cr
         1&1&0&0&0&0&0&0}\sigmad\nonumber
\eea

In total, we find twelve distinct cylinder partition
functions~\cite{Cardy89,AfflOS98}
\begin{eqnarray}
Z_{A|A}(q)&=&\chit_{1,1}(q)=\chi_{1,1}(q)+\chi_{1,5}(q)\nonumber\\
Z_{A|B}(q)&=&\chit_{1,4}(q)=\chi_{1,3}(q)\nonumber\\
Z_{A|AB}(q)&=&\chit_{2,4}(q)=\chi_{3,3}(q)\nonumber\\
Z_{A|BC}(q)&=&\chit_{2,1}(q)=\chi_{3,5}(q)+\chi_{3,1}(q)\nonumber\\
Z_{A|F}(q)&=&\chit_{1,2}(q)=\chi_{4,2}(q)+\chi_{4,4}(q)\nonumber\\
Z_{A|N}(q)&=&\chit_{2,2}(q)=\chi_{2,2}(q)+\chi_{2,4}(q)\nonumber\\
&=&Z_{AB|F}(q)\nonumber\\
Z_{AB|AB}(q)&=&\chit_{1,1}(q)+\chit_{2,1}(q)\nonumber\\
&=&\chi_{1,1}(q)+\chi_{3,5}(q)+\chi_{3,1}(q)+\chi_{1,5}(q)\nonumber\\
Z_{AB|AC}(q)&=&\chit_{1,4}(q)+\chit_{2,3}(q)=\chi_{3,3}(q)+\chi_{1,3}(q)\nonumber\\
Z_{AB|N}(q)&=&\chit_{2,2}(q)+\chit_{1,2}(q)\nonumber\\
&=&\chi_{2,2}(q)+\chi_{2,4}(q)+\chi_{4,2}(q)+\chi_{4,4}(q)\nonumber\\
Z_{F|F}(q)&=&\chit_{1,1}(q)+\chit_{1,3}(q)+\chit_{1,4}(q)\nonumber\\
&=&\chi_{1,1}(q)+\chi_{1,5}(q)+2\chi_{1,3}(q)\nonumber\\
Z_{F|N}(q)&=&\chit_{2,1}(q)+\chit_{2,3}(q)+\chit_{2,4}(q)\nonumber\\
&=&\chi_{3,5}(q)+\chi_{3,1}(q)+2\chi_{3,3}(q)\nonumber\\
Z_{N|N}(q)&=&\chit_{1,1}(q)+\chit_{2,1}(q)+\chit_{1,3}(q)\nonumber\\
&&\quad\mbox{}+\chit_{1,4}(q)+\chit_{2,3}(q)+\chit_{2,4}(q)\nonumber\\
&=&\chi_{1,1}(q)+\chi_{3,5}(q)+\chi_{3,1}(q)+\chi_{1,5}(q)\nonumber\\
&&\quad\mbox{}+2\chi_{3,3}(q)+2\chi_{1,3}(q)\nonumber
\end{eqnarray}
\goodbreak
Here we restrict to Virasoro characters with $r+s$ even.
The symmetry
\be
Z_{(r_1,a_1)|(r_2,a_2)}(q)=Z_{(r_2,a_2)|(r_1,a_1)}(q)
\ee
follows because the characters do not distinguish between a field $\phit$
and its conjugate
$\phit^\dagger$.

\section{Integrable Boundary Weights}

\nc{\ru}[1]{\rule[-#1ex]{0ex}{#1ex}}
\nc{\B}[5]{\setlength{\unitlength}{4mm}
\begin{array}{@{}c@{}}
\begin{picture}(2.2,2.2)(-0.1,0.16)
\multiput(1.6,0.1)(0,0.3){7}{\line(0,1){0.2}}
\put(0.6,1.1){\line(1,-1){1}}\put(0.6,1.1){\line(1,1){1}}
\put(1.8,2.2){\spos{tl}{#5}}\put(0.45,1.1){\spos{r}{#4}}
\put(1.8,0){\spos{bl}{#3}}\end{picture}\end{array}}

We conjecture that conformal boundary conditions for $sl(2)$ models can be
realized as integrable boundary conditions for the associated lattice
models~\cite{ABF84}. For the
$(A_{g-1},A_g)$ theories the integrable boundary weights have been
obtained~\cite{BehrP96}, as solutions to the boundary Yang-Baxter equation,
by a fusion construction.
This method generalizes~\cite{BehrP96} to the \ade\ models using the
appropriate fusion
process~\cite{ZhouP94}.
The solutions to the boundary Yang-Baxter
equation are naturally labelled by a pair $(r,a)$
and are constructed
by starting at $a$ and fusing $r-1$ times.
For $(A_4,D_4)$
, the non-zero boundary weights
are given
explicitly by
\begin{eqnarray*}
\ru{3.5}A,B,C&\!=\!&(1,a):\B{1}{1}{a}{2}{a}=1,\quad a=1,3,4\\
\ru{3.5}F&\!=\!&(1,2):\B{1}{2}{2}{1}{2}=\!\B{1}{2}{2}{3}{2}=
\B{1}{2}{2}{4}{2}=1\\
\ru{3.5}BC&\!=\!&(2,1):\B{2}{1}{2}{3}{2}=\!\B{2}{1}{2}{4}{2}=\rho_1(u)\;,\;
\B{2}{1}{2}{1}{2}=\rho_1(-u)\\
\ru{3.5}AC&\!=\!&(2,3):\B{2}{3}{2}{1}{2}=\!\B{2}{3}{2}{4}{2}=\rho_1(u)\;,\;
\B{2}{3}{2}{3}{2}=\rho_1(-u)\\
\ru{4}AB&\!=\!&(2,4):\B{2}{4}{2}{1}{2}=\!\B{2}{4}{2}{3}{2}=\rho_1(u)\;,\;
\B{2}{4}{2}{4}{2}=\rho_1(-u)\\
N&\!=\!&(2,2):\left\{\begin{array}{@{\:}l@{}}
\ru{3.5}
\B{2}{2}{a}{2}{b}
=\rho_2(u),\quad a\ne b,\quad a,b=1,3,4\\
\B{2}{2}{a}{2}{a}=
\rho_3(u),\quad a=1,3,4
\end{array}
\right.
\end{eqnarray*}
with $u$ the spectral parameter, $\lambda=\pi/6$, $\xi$ arbitrary and
$$
\begin{array}{@{}l@{}}
\ru{3}\displaystyle\rho_1(u)\!=\!
\frac{\sin(u\!-\!\lambda\!-\!\xi)\:
\sin(u\!-\!\lambda\!+\!\xi)}{\sin^2\!\lambda},\qquad
\rho_2(u)\!=\!
\frac{\sin\!2u}{\sin\!2\lambda}\\
\ru{3}\displaystyle\rho_3(u)\!=\!
\frac{2\sin(u\!-\!\xi)\sin(u\!+\!\xi)\!+\!
\sin(u\!-\!2\lambda\!-\!\xi)
\sin(u\!-\!2\lambda\!+\!\xi)}{\sin^2\!2\lambda}.
\end{array}
$$
The new boundary condition~\cite{AfflOS98} $N$ is found to be
antiferromagnetic in nature. The value of $u$ should be set to its
isotropic value $u=\lambda/2$ and $\xi$ chosen appropriately
to obtain the conformal boundary conditions.

\section{Conclusion}

In conclusion
we have proposed a set of conjectures that extend
the theory of conformal boundaries in a consistent way.
The structure of the partition functions is dictated
by a new fusion algebra.
We comment that the conjecture  (\ref{CPFs}c)
is independent of the choice of endpoint node and
eigenvectors and is
meaningful for $D_{2n+1}$  and $E_7$, even though
a proper understanding of the  fusion matrices in (2.4b)
is missing.  We expect the
 extension to higher rank~\cite{DFZLZZ} to be straightforward.
A much more comprehensive
version of this work will be published elsewhere.


\begin{references}
\bibitem{Cardy89}J. L. Cardy, Nucl.\  Phys.\ {\bf B324}, 581 (1989).
\bibitem{Verlinde88}E. Verlinde, Nucl.\  Phys.\ {\bf B300}, 360 (1988).
\bibitem{AfflOS98}I. Affleck, M. Oshikawa and H. Saleur, Boundary Critical
Phenomena in the Three-State Potts Model, cond-mat/9804117.
\bibitem{FuchsS98}J. Fuchs and C. Schweigert, Completeness of Boundary
Conditions for the Critical Three-State Potts Model,
hep-th/9806121.
\bibitem{CapItzZub87}A. Cappelli, C. Itzykson and J.-B. Zuber,
Comm.\ Math.\ Phys.\ {\bf 113}, 1 (1987).
\bibitem{Pas}V. Pasquier, {Mod\`eles Exacts Invariants Conformes},
Th\`ese d'Etat, Orsay, 1988.
\bibitem{DFZPZ}P. Di Francesco, and J.-B. Zuber, in Recent Developments in
Conformal Field Theories, eds.
S. Randjbar-Daemi, E. Sezgin and J.-B. Zuber, World Scientific (1990); P.
Di Francesco, Int.\ J. Mod.\
Phys.\ {\bf A7}, 407 (1992).
\bibitem{SaleurB89}H. Saleur and M. Bauer, Nucl.\ Phys.\ {\bf B320}, 591
(1989).
\bibitem{PasqS90}V. Pasquier and H. Saleur, Nucl.\ Phys.\ {\bf B330}, 523
(1990).
\bibitem{ABF84}G. E. Andrews, R. J. Baxter and P. J. Forrester, J. Stat.\
Phys.\ {\bf 35}, 193 (1984);
P. J. Forrester and R. J. Baxter, J. Stat.\ Phys.\ {\bf 38}, 435 (1985);
V. Pasquier, Nucl.\ Phys.\ {\bf B285}, 162 (1987).
\bibitem{BehrP96}R. E. Behrend, P. A. Pearce and D. L. O'Brien, J. Stat.\
Phys.\ {\bf 84}, 1 (1996);
R. E. Behrend and P. A. Pearce, J. Phys.\ {\bf A29}, 7827 (1996);
R. E. Behrend and P. A. Pearce, Int.\ J. Mod.\ Phys.\ {\bf B11}, 2833 (1997).
\bibitem{ZhouP94}Y.-K. Zhou and P. A. Pearce, Int.\ J. Mod.\ Phys.\ {\bf
B8}, 3531 (1994).
\bibitem{DFZLZZ}P. Di Francesco, and J.-B. Zuber, Nucl.\ Phys.\ {\bf B338},
602 (1990);
S. Loesch, Y.-K. Zhou and J.-B. Zuber, Int.\ J. Mod.\ Phys.\ {\bf A12}, 4425
(1997).

\end{references}
\end{document}